\documentclass[iop,apj,tighten]{emulateapj}
\usepackage{mathtools, graphicx, booktabs, apjfonts}
\usepackage[usenames,dvipsnames]{color}
\usepackage{amssymb,amsmath}
\usepackage{multirow}
\usepackage{upquote}

\shorttitle{Early Venus Obliquity Variations}
\shortauthors{Barnes, Quarles, Lissauer, Chambers, \& Hedman}

\begin{document}

\title{Obliquity Variability of a Potentially Habitable Early Venus}

\author{Jason W. Barnes\altaffilmark{1},Billy Quarles\altaffilmark{2,3}, 
Jack J. Lissauer\altaffilmark{2},John Chambers\altaffilmark{4},Matthew M.
Hedman\altaffilmark{1}}
\altaffiltext{1}{Department of Physics; University of Idaho; Moscow, ID 
83844-0903, USA; ResearcherID:  B-1284-2009; email: {\tt jwbarnes@uidaho.edu}}
\altaffiltext{2}{Space Science and Astrobiology Division MS 245-3; NASA Ames Research Center; Moffett Field, CA 94035, USA}
\altaffiltext{2}{Department of Physics and Physical Science, The University of
Nebraska at Kearney, Kearney, NE 68849, USA}
\altaffiltext{4}{Department of Terrestrial Magnetism; Carnegie Institution of Washington; 
Washington, DC 20015-1305, USA}

\begin{abstract} 

Venus currently rotates slowly, with its spin controlled by solid-body  and
atmospheric thermal tides. However, conditions may have been far  different 4
billion years ago, when the Sun was fainter and most of the carbon  within Venus
could have been in solid form, implying a low-mass  atmosphere. We investigate
how the obliquity would have varied for a  hypothetical rapidly rotating Early
Venus. The obliquity variation  structure of an ensemble of hypothetical Early
Venuses is simpler than that Earth would have if it lacked its large Moon
\citep{moonless.Earth}, having just one  primary chaotic regime at high prograde
obliquities.  We note  an unexpected long-term variability of up to $\pm7^\circ$
for retrograde  Venuses. Low-obliquity Venuses show very low total obliquity
variability  over billion-year timescales -- comparable to that of the real 
Moon-influenced Earth.

\end{abstract}


\section{Introduction}

The obliquity $\Psi$ --- defined as the angle between a planet's rotational
angular momentum and its orbital angular momentum --- is a fundamental dynamical
property of a planet.  A planet's obliquity influences its climate and potential
habitability.  Varying orbital inclinations and precession of the orbit's
ascending node can alter obliquity, as can torques exerted upon a planet's
equatorial bulge by other planets.  The Earth exhibits a relatively stable and
benign long-term climate because our planet's obliquity varies only of order
$\sim3^\circ$.  As a point of comparison, the obliquity of Mars varies over a
very large range --- $\sim0^\circ$--$60^\circ$
\citep{1993Sci...259.1294T,1993Natur.361..615L,2004Icar..170..343L}. 

Changes in obliquity drive changes in planetary climate.  In the case where
those obliquity changes are rapid and/or large, the resulting climate shifts can
be commensurately severe \citep[see, e.g.,][]{2004Icar..171..255A}.  The Earth's
present climate resides at a tipping point between glaciated and non-glaciated
states, and the small $\sim3^\circ$ changes in our obliquity from Milankovic
Cycles drives glaciation and deglaciation of northern Europe, Siberia, and North
America \citep{milankovivc1998canon}.  These glacial/interglacial cycles reduce
biodiversity in periodically glaciated Arctic regions
\citep[e.g.,][]{araujo2008quaternary,hawkins2003relative,hortal2011ice}.  The
resulting insolation shifts jolt climatic patterns worldwide, causing species in
affected regions to migrate, adapt, or be rendered extinct.

Perhaps paradoxically, large-amplitude obliquity variations can also act 
to favor a planet's overall habitability.  Low values of obliquity can 
initiate polar glaciations that can, in the right conditions, expand 
equatorward to envelop an entire planet like the ill-fated ice-planet 
Hoth in \emph{The Empire Strikes Back} \citep{george1980empire}.  
Indeed, our own planet has experienced so-called Snowball Earth states 
multiple times in its history \citep{1998Sci...281.1342H}.  Although 
high obliquity drives severe seasonal variations, the annual average 
flux at each surface point is more uniform on a high-obliquity world 
than the equivalent low-obliquity one.  Hence high obliquity can act to 
stave off snowball states \citep{2015E&PSL.415..100S} and extreme 
obliquity variations may act to expand the outer edge of the habitable 
zone \citep{2014AsBio..14..277A} by preventing permanent snowball 
states.

Thus knowledge of a planet's obliquity variations may be critical to the
evaluation of whether or not that planet provides a long-term habitable
environment.  A planet's siblings affect its obliquity evolution primarily via
nodal precession of the planet's orbit.  Obliquity variations become chaotic
when the precession period of the planet's rotational axis (26,000 years for
Earth) becomes commensurate with the nodal precession period of the planet's
orbit ($\sim100,000$ years for Earth).  Secular resonances, those that only
involve orbit-averaged parameters as opposed to mean-motion resonances for which
the orbital periods are near-commensurate, typically cluster together in the
Solar System such that if you are near one secular period, you are likely near
others as well.  And those clusters of secular resonances act to drive chaos
that increases the range of a planet's obliquity variations.  

The gravitational influence of Earth's Moon speeds the precession of our
rotation axis, and stabilizes our obliquity.  Without this influence Earth's
rotation axis precession would have a period of $\sim100,000$ years, close
enough to commensurability as to drive large and chaotic obliquity variability
\citep{1993Natur.361..615L}.  Though our previous work \citep{moonless.Earth}
showed that such variations would not be as large as those of Mars, the
difference between commensurate precessions and non-commensurate precessions is
stark.

\citet{2007Icar..188....1A} investigated the obliquity evolution of potentially
habitable extrasolar planets with large moons, following on work by
\citep{2004Icar..168..223A} showing the generalized influence of nearby giant
planets on terrestrial planet obliquity in general.  \citet{2014MNRAS.440.3685B}
studied the obliquity variations in for the specific super-Earth HD40307g.

To expand the general understanding of potentially habitable worlds' 
obliquity variations, we use the only planetary system that we know well 
enough to render our calculations accurate: our own.  In this paper, we 
analyze the obliquity variations of a hypothetical Early Venus as an 
analog for potentially habitable exoplanets.

Venus \emph{was} likely in the Sun's habitable zone 4.5~Gyr ago, when the Sun
was only 70\% its present luminosity  \citep{1993ApJ...418..457S}.  Such an
Early Venus could well have had a low-mass atmosphere (with most of the planet's
carbon residing within rocks), and tides would not yet have substantially damped
its spin rate.  In fact, \citet{2011AsBio..11..443A} suggest that the real Venus
may have been habitable as recently as 1~Gyr ago, provided that its initial
water content was small (as might result from impact-driven dessication, as per
\citet{2015E&PSL.429..181K}, or because the planet is located well interior to
the ice line).

\begin{table*}[bth]
\begin{center}

\begin{tabular}{l|cc|cc|cc}
\hline 
\hline
Rotation  & \multicolumn{2}{c|}{LR93a} & \multicolumn{2}{c|}{CLS03} & \multicolumn{2}{c}{\citet{moonless.Earth}}  \\
Period (hr) &  $J_2$    &  (\textquotesingle\textquotesingle\ /yr) &  $J_2$    &  (\textquotesingle \textquotesingle\ /yr) &  J$_2$    &  (\textquotesingle \textquotesingle\ /yr)\\
\hline
   4 &	1.405422e-02 &	99.94621 &	4.694002e-02 &	334.75028 & 4.692702e-02 &	334.63436 \\
   8 &	3.504570e-03 &	49.84532 &	1.036030e-02 &	147.76787 & 1.034730e-02 &	147.57222 \\
  12 &	1.555255e-03 &	33.18048 &	4.526853e-03 &	96.84904 & 4.513853e-03 &	96.56422 \\
  16 &	8.739020e-04 &	24.85893 &	2.536138e-03 &	72.34533 & 2.523138e-03 &	71.96950 \\
  20 &	5.588364e-04 &	19.87076 &	1.623182e-03 &	57.87820 & 1.610182e-03 &	57.41067 \\
  24 &	3.878196e-04 &	16.54781 &	1.129453e-03 &	48.32779 & 1.116453e-03 &	47.76823 \\
  30 &	2.480000e-04 &	13.22734 &	7.266291e-04 &	38.86438 & 7.136291e-04 &	38.16642 \\
  36 &	1.721063e-04 &	11.01537 &	5.082374e-04 &	32.62021 & 4.952374e-04 &	31.78363 \\
\hline

\end{tabular}
\caption{Values for the zonal harmonic ($J_2$) and `precession' constant
($\alpha$) determined from the rotation period for the models presented in
\citet{1993Natur.361..608L}, \citet{2003Icar..163....1C}, and
\citet{moonless.Earth}.}
	\label{table:IC}
\end{center}
\end{table*}

\begin{figure*}[!bth]
\centering
\includegraphics[width=\linewidth]{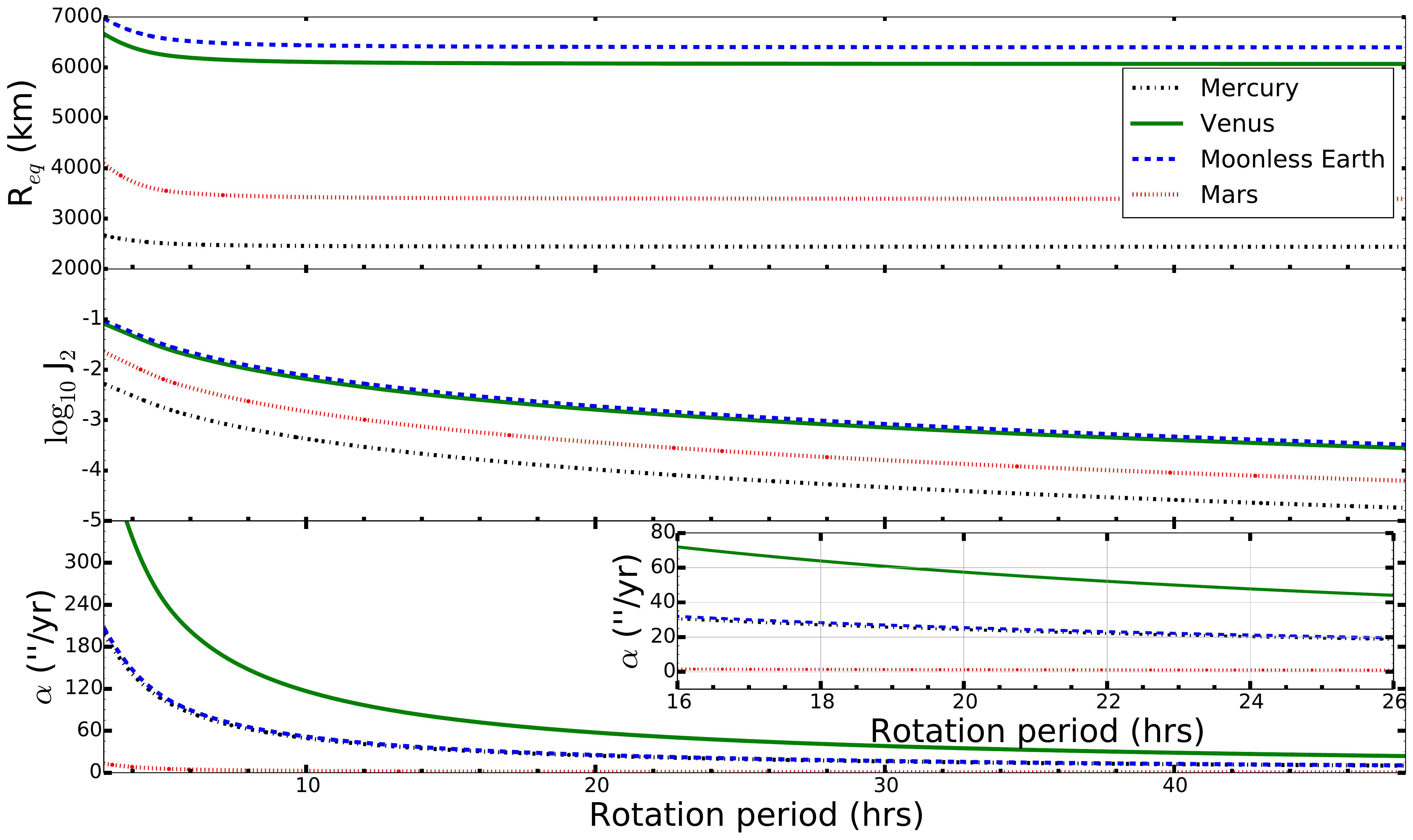}
\caption{Illustration of how the derived starting values for the equatorial
radius due to rotation (R$_{eq}$), zonal harmonic (J$_2$), and `precession'
constant ($\alpha$) vary in response to the initial rotation period from 4 to 48
hrs using the formalism given by \citet{moonless.Earth}.  Curves are provided
considering each of the terrestrial planets (Mercury, Venus, Moonless 
Earth (a 
single
planet with the mass of the Moon added to that of the Earth), and Mars).}
\label{figure:Spin_param}
\end{figure*} 

In this work we numerically explore the obliquity variations of Early Venus with
a parameter grid study that incorporates a wide variety of rotation rates and
obliquities.  Note that this work is \emph{not} intended to study Venus'
\emph{actual} historical obliquity state, information about which has been
destroyed by its present tidal equilibrium
\citep{2003Icar..163....1C,2003Icar..163...24C}.  Instead, we use Venus with a
wide range of assigned rotation rates and initial obliquities as an analog for
habitable exoplanets and to explore what types of obliquity behavior were
possible for a potentially habitable Early Venus.  Our methods build on those of
\citet{moonless.Earth} and are described in Section 2.  We provide qualitative
and quantitative descriptions of the drives of obliquity variations and chaos in
Section 3.  Results of our simulations are presented in Section 4, and we
conclude in Section 5.  Readers interested primarily in the results might
consider jumping to Section 4, while those also interested in the physics of why
obliquity varies can add Section 3.

\section{Methodology} \label{section:methodology}

\subsection{Approach}

We track the evolution of obliquity for the hypothetical Venus computationally,
using a modified version of the mixed-variable sympletic (MVS) integration
algorithm within the \texttt{mercury} package developed by
\cite{1999MNRAS.304..793C}.  The modified algorithm \texttt{smercury} (for
spin-tracking \texttt{mercury}) explicitly calculates both orbital forcing for
the 8-planet Solar System and spin torques on one particular planet in the system
from the Sun and sibling planets following \citet{1994AJ....107.1189T}.  Our
explicit numerical integrations represent an approach distinct from the
frequency-mapping treatment employed by \citet{1993Natur.361..615L}.  See
\citet{moonless.Earth} for a complete mathematical description of our
computational techniques. 

The \texttt{smercury} algorithm treats the putative Venus as an axisymmetric
body.  In so doing, we neglect both gravitational and atmospheric tides.  Tidal
influence critically drives the present-day rotation state of real Venus
\citep{2001Natur.411..767C}.  We are interested in an early stage of dynamical
evolution, however, where the tidal effects do not dominate.  Therefore we
consider only solar and interplantary torques on the rotational bulge.  The
simultaneous consideration of tidal and dynamical effects is outside the scope
of the present work.

In the case of differing rotation periods, we incorporate the planet's dynamical
oblateness and its effects on the planet's gravitational field.  These effects
manifest as the planet's gravitational coefficient $J_2$, values for which we
determine from the Darwin-Radau relation, following Appendix A of
\citet{moonless.Earth}.   Additionally, as in \citet{moonless.Earth} we employ
``ghost planets'' to increase the efficiency of our calculations --- essentially
we calculate planetary orbits just one time, while assuming a variety of
different hypothetical Venuses for which we calculate just the obliquity
variations.   We neglect (the very small effects of) general relativity and
stellar $J_2$.

\subsection{Initial Conditions}

We select orbital initial conditions with respect to the J2000 epoch  where the
Earth-Moon barycenter resides coplanar with the ecliptic.  
\citet{2014ApJ...795...67L} and \cite{2011Icar..213..423B} investigated how
obliquity variations are affected by alternate early Solar System
architectures, specifically the Nice  model \citep{2007AJ....134.1790M}. As an
investigation of the long-term  characteristic obliquity behavior, however, we
instead elect to  integrate the present Solar System orbits, which are known to
much  higher accuracy.

\citet{moonless.Earth} showed that chaotic variations in obliquity for a 
Moonless Earth can manifest from slightly different initial orbits.  We 
thus remove this effect by using a common orbital solution for all of 
our simulations.  We assume the density of our hypothetical Venus to be 
the same as the real Venus, 5.204 g/cm$^3$.  However, we assume a moment 
of inertia coefficient to be the same as that for the real Earth 
\citep[0.3296108;][]{GlobalEarthPhysics} given that a truly habitable 
Venus would likely have a different internal structure than the real 
one.  We do not vary the moment of inertia with rotation period.

We consider a range of rotation periods of between 4 and 36 hours.  The 
short end is set by the rotation speed at which the planet would be near 
breakup, where our Darwin-Radau and axisymmetric assumptions break down.  
The longer limit represents a value 50\% longer than Earth's rotation, 
which itself has been tidally slowed over the past 4.5~Gyr.  In the epoch 
of Solar System history that we consider, Earth's own day was 
significantly shorter than it is today. 


Along with various rotation rates, we also consider initial obliquity values
$\Psi$ that range from $0^\circ$ to $180^\circ$.  Planets with obliquity
between $90^\circ$ and $180^\circ$ rotate retrograde to their orbital motions.  
Obliquity alone does not completely determine the orientation of a planet's spin
axis in space (unless $\Psi=0^\circ$ or $\Psi=180^\circ$).  Therefore, for each
obliquity we also consider various initial axis azimuths, $\varphi$, which
correspond to the direction that the spin pole points.  

In order to generate the proper initial spin states, we define the angles of
obliquity and azimuth.  \citet{moonless.Earth} used a similar approach; however,
that previous study was for the Earth-Moon barycenter with zero initial
inclination relative to the ecliptic plane.  In contrast, the definition of spin
direction for any other planet requires two additional rotations that include
that planet's inclination, $i$, and its ascending node, $\Omega$, both relative
to the J2000 ecliptic.  Thus the general rotation matrices $R_1$ and $R_2$ can be used
to define the desired obliquity, $\Psi$, and azimuth, $\varphi$:
\begin{equation}
R_1(\gamma) = \begin{pmatrix}
1 & 0 & 0\\
0 & \cos \gamma & -\sin \gamma \\
0 & \sin \gamma & \cos \gamma 
\end{pmatrix}
{\rm and \;\;}
R_2(\beta) = \begin{pmatrix}
\cos \beta & \sin \beta & 0\\
-\sin \beta & \cos \beta & 0\\
0 & 0 & 1 \\
\end{pmatrix} 
{\rm ~~~~~.}
\end{equation}  
The azimuthal angles, $\varphi$ and $\Omega$, undergo rotations via $R_2$ with
$\beta=\varphi$ or $\beta=\Omega$, where the altitudinal angles are rotated
using $R_1$ with $\gamma=\Psi$ and $\gamma=i$.

\section{Obliquity Evolution}

A planet's obliquity, $\Psi$, is defined as the magnitude of the angular
distance between the direction of the angular momentum vector for a planet's
spin and that for its orbit.  Therefore obliquity can change if either of those
two vectors change direction:  (1) the rotational angular momentum vector or (2)
the orbital angular momentum vector.  Let's consider each in turn.

\begin{figure}[tbh]
\centering
\includegraphics[width=\columnwidth]{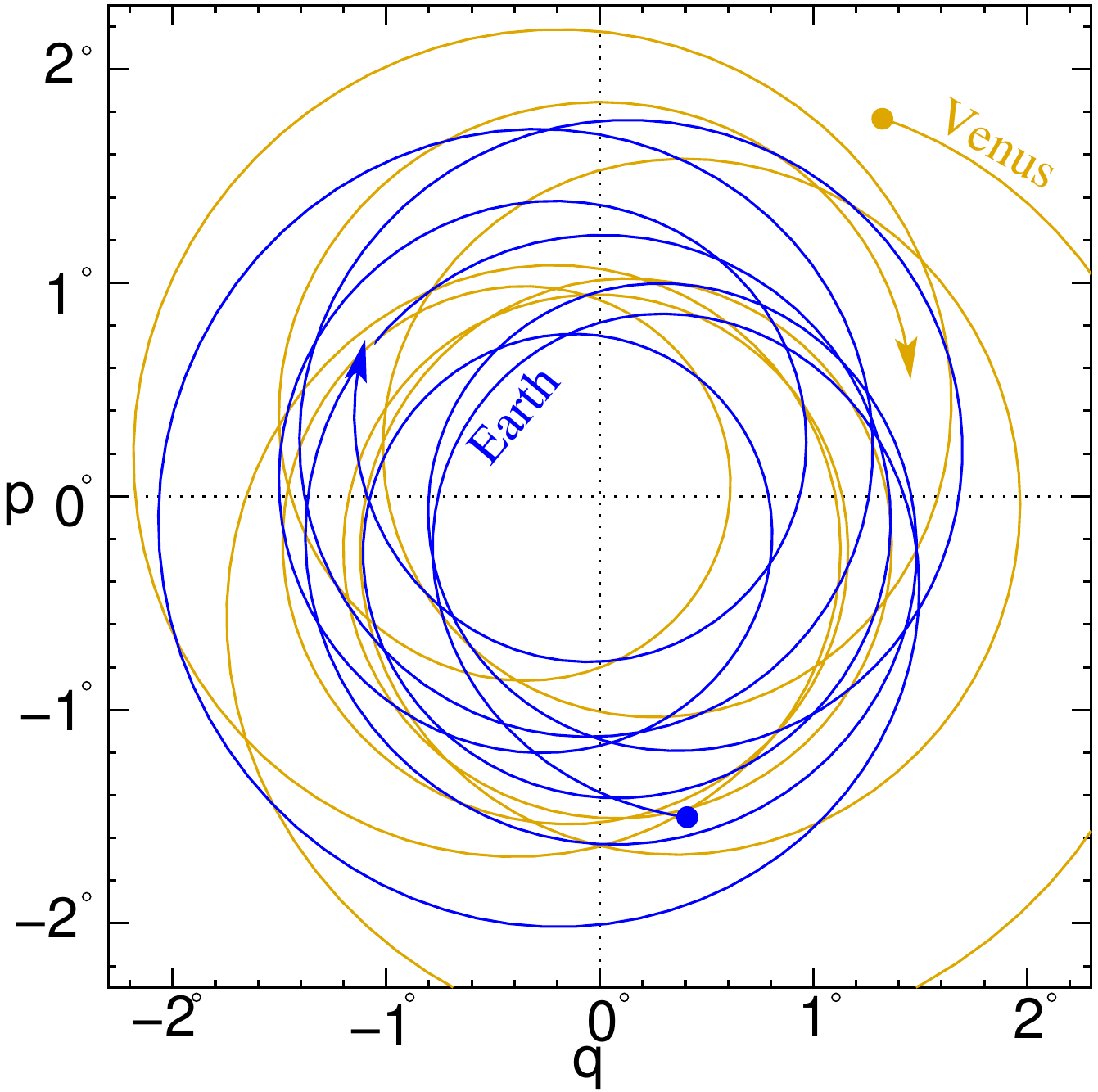}
\caption{This plot shows the projected direction in which Venus' (yellow) and
Earth's (blue) orbital angular momentum points as it varies over the course of 500,000 years from the
present day.  The projection is in $p$-$q$ space, with $p\equiv I\sin(\Omega)$ and
$q\equiv I\cos(\Omega)$ where $\Omega$ is the longitude of the ascending node of the
orbit and $I$ is the orbital inclination.  The indicated motion represents nodal
precession, where a planet's orbit reorients in space like a coin spinning
down on a desktop.}
\label{figure:VE_pq}
\end{figure}

\subsection{Rotational Angular Momentum}

Torques on a planet's rotational bulge from the Sun primarily drive changes in
the direction of that planet's rotational axis.  Because the star must always be
located within the plane of the planet's orbit, however, these changes
\emph{cannot} directly alter the planet's obliquity $\Psi$.  Instead, the
stellar torque induces the planetary rotation axis to precess around the orbit
normal, constantly changing the axis azimuth $\varphi$, but leaving the
obliquity $\Psi$ unchanged.  This effect is called the precession of the
equinoxes.  It is why the date of the equinox slowly creeps forward over time,
and why Polaris has not always been near the Earth's north pole \citep[see,
e.g.,][]{karttunen2007fundamental}.
 
\begin{figure}[tbh]
\centering
\includegraphics[width=\columnwidth]{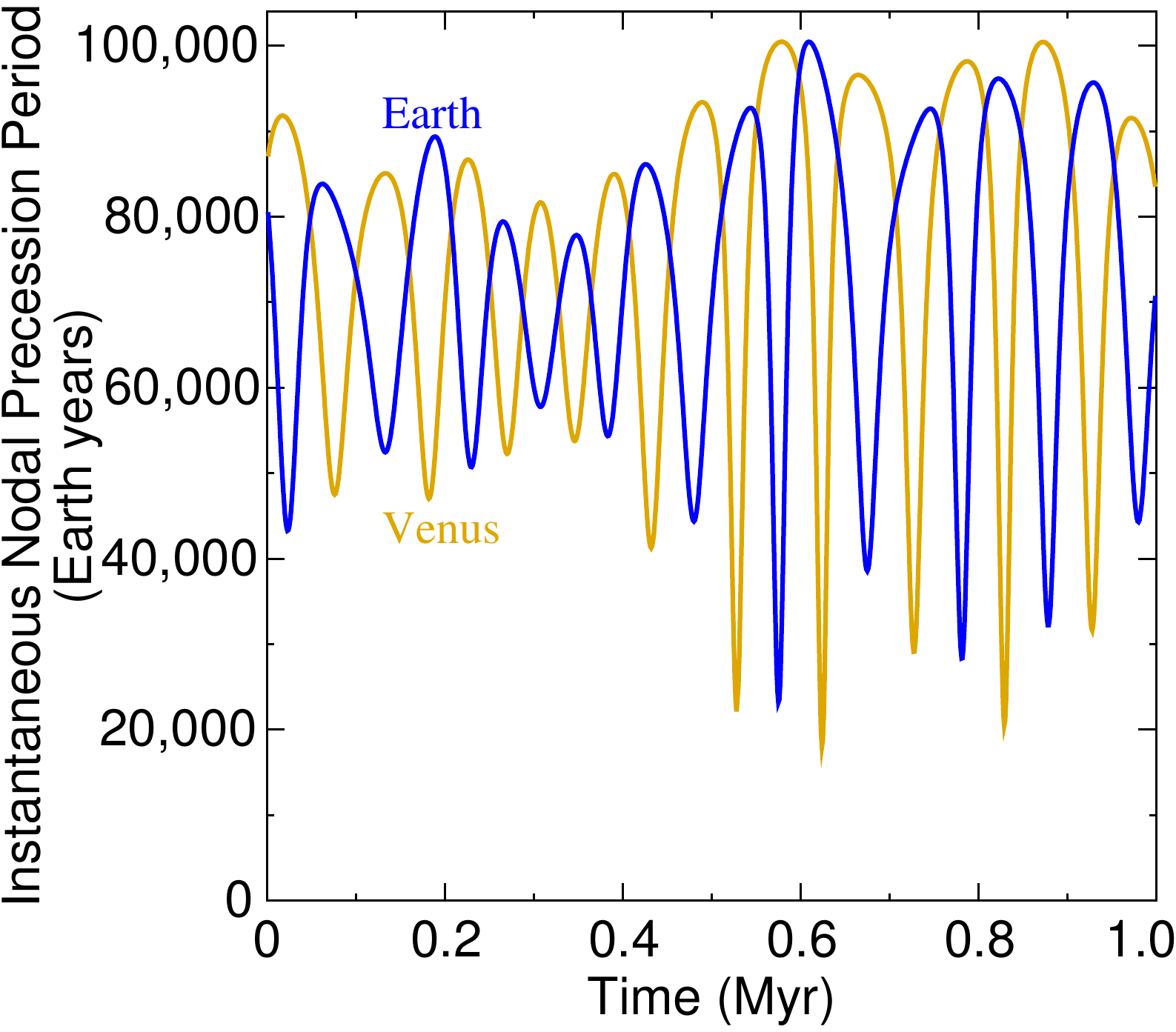}
\caption{While Figure \ref{figure:VE_pq} shows the \emph{direction} that the
orbit poles of Venus and Earth point to, it lacks a timescale.  This plot
provides such a timescale, as it depicts the instantaneous precession period
(i.e., $2\pi/\frac{\mathrm{d}\varphi}{\mathrm{d}t}$ for both Venus (yellow) and
Earth (blue) over a million years.  Because Venus and Earth each represent the
primary influence on the nodal precession of the other and because they are of
comparable mass, the long-term average precession rates for the two are about
the same at $\sim70,000$~years.} 
\label{figure:VE_nodal_period}
\end{figure}

\begin{figure*}[tbh]
\centering
\includegraphics[width=\linewidth]{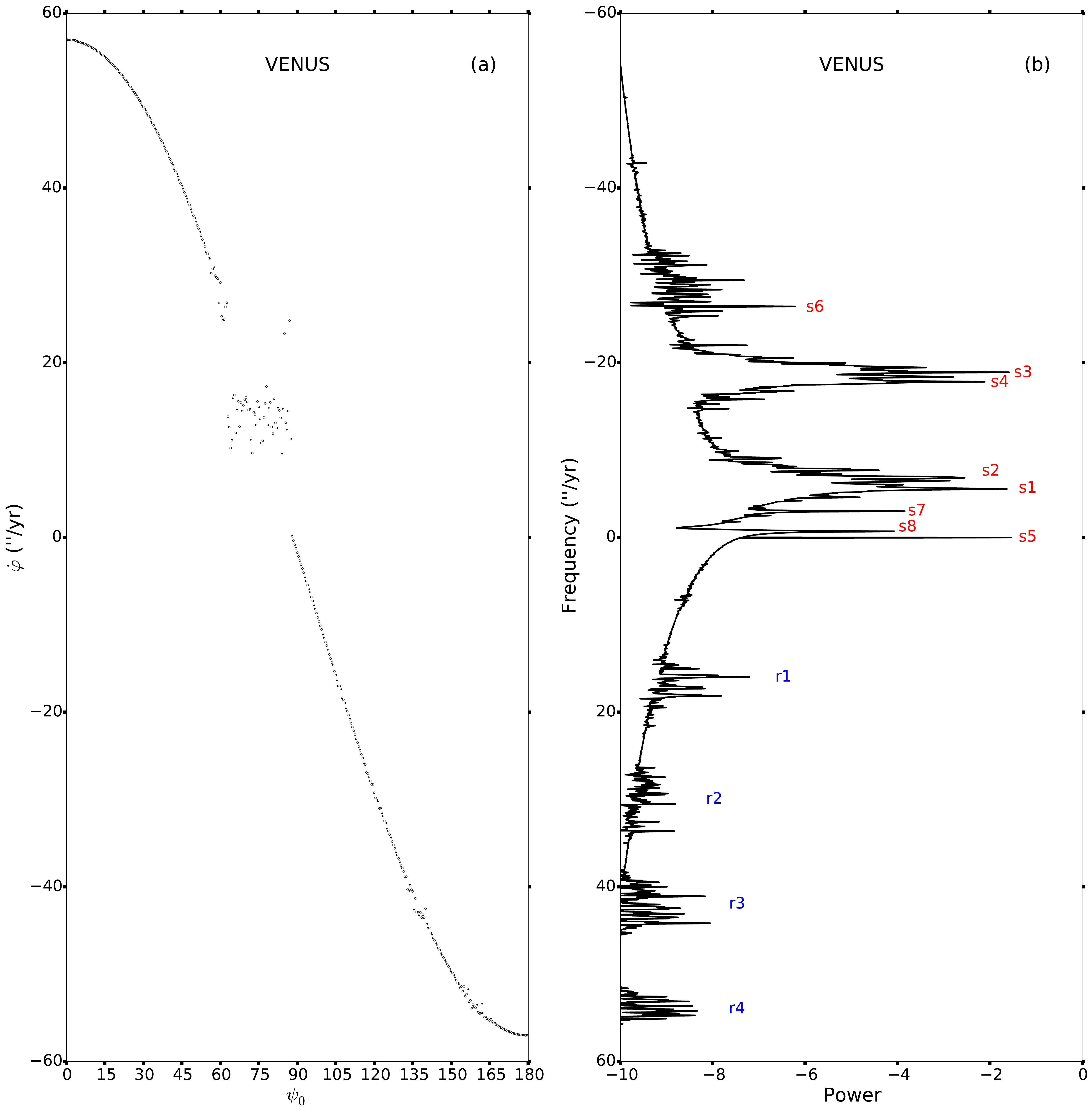}
\caption{Precession frequencies (a) for a hypothetical Venus with a rotation
period of 20 hours.  Panel (a) shows the average precession rate $\dot{\varphi}$
as a function of initial obliquity $\Psi_0$.  Chaotic zones appear
for obliquities ranging from 60$^\circ-90^\circ$ that correlates to the
precession frequencies ranging from 0 -- 
26\textquotesingle\textquotesingle
/yr showing correspondence with the main secular orbital frequencies~(b) of the
Solar System \citep{1993Natur.361..608L,1996CeMDA..64..115L}.  The power 
shown
in the $x$-axis of panel (b) is logarithmic.}
\label{figure:precession}
\end{figure*} 

\begin{figure*}[tbh]
\centering
\includegraphics[width=\linewidth]{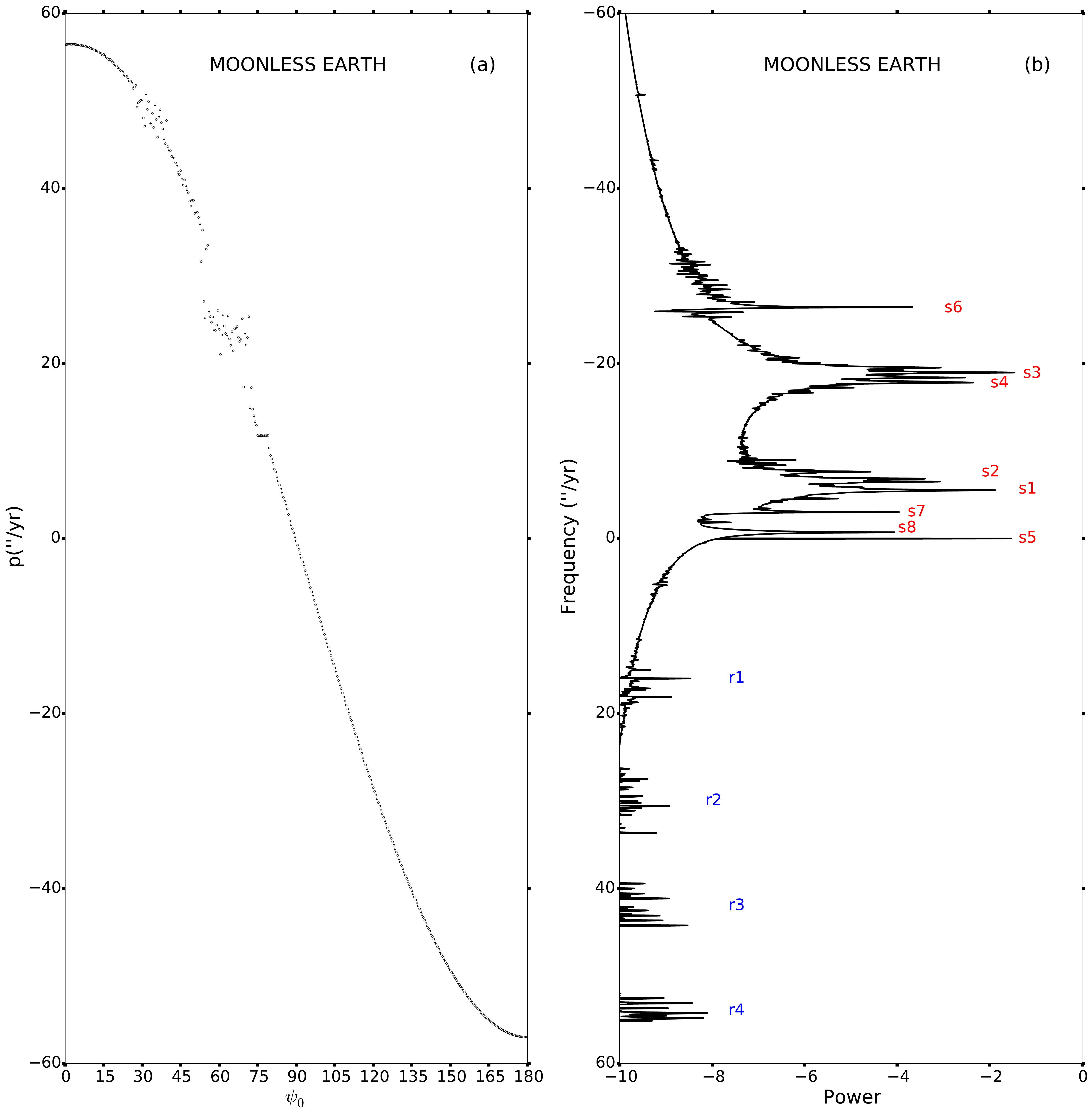}
\caption{Similar to Figure \ref{figure:precession}, here we show the precession
frequencies (a) and the power spectrum of the orbital angular momentum vector
(b) for a 24-hour-rotation Moonless Earth for comparison with our hypothetical
Venuses.}
\label{figure:precession_earthmoo}
\end{figure*} 

The rate of axial precession depends on the planet's dynamical oblateness
gravitational coefficient $J_2$, the mass and
distance from the Sun, and the planet's moment of inertia.  The rate also depends
weakly on the obliquity itself; therefore precession rates are typically given
in terms of the precession constant $\alpha$, where
\begin{equation}
\dot\varphi=\alpha\cos(\Psi)~~~~.
\label{eq:precession_rate}
\end{equation}
In Table \ref{table:IC} we show the values of Venus' zonal harmonic ($J_2$) and
precession constant ($\alpha$) for both the present study and previous work for
a range of rotation periods.  Our values strongly resemble  those of
\cite{2003Icar..163....1C} but differ substantially from those used by
\cite{1993Natur.361..608L}\footnote{\cite{1993Natur.361..608L} provides a
formalism to derive the value of $\alpha$, but does not indicate a precise
determination of the equatorial flattening ($C-A \over C$).  In order to
determine the appropriate starting values, we produce a power law fit using
their Figure 5b.  From this power law, the initial values of $\alpha$ (and hence
$J_2$) are reduced by a factor of $\sim3$.}.  Figure \ref{figure:Spin_param}
grapically represents the equatorial radius, $J_2$, and precession constant
$\alpha$ for our hypothetical Early Venuses as a function of their rotation
period.

In general, axial precession for Venus occurs about twice as fast as axial
precession for an equivalent planet at 1~AU.  Because the Sun's gravity drives
axial precession, the fact that Venus' semimajor axis is nearly $\sqrt{2}$~AU
explains the factor of 2 faster axial precession.  Functionally, for obliquity
variations, the Sun speeds Venus' axial precession in a similar manner that the
Moon speeds Earth's.  

An expectation might be that Early Venus' obliquity variations should more
closely resemble that of real-life Earth with the Moon than that of the moonless
Earth from \citet{moonless.Earth}.  Circumstances that act to slow Venus' axial
precession from that of the precession constant --- such as a smaller rotational
bulge, or a higher obliquity --- could act to bring the axial precession rate
into near-commensurability with precession rates of the orbital ascending node,
leading to chaotic obliquity evolution.

\begin{figure*}[tbh]
\centering
\includegraphics[width=\linewidth]{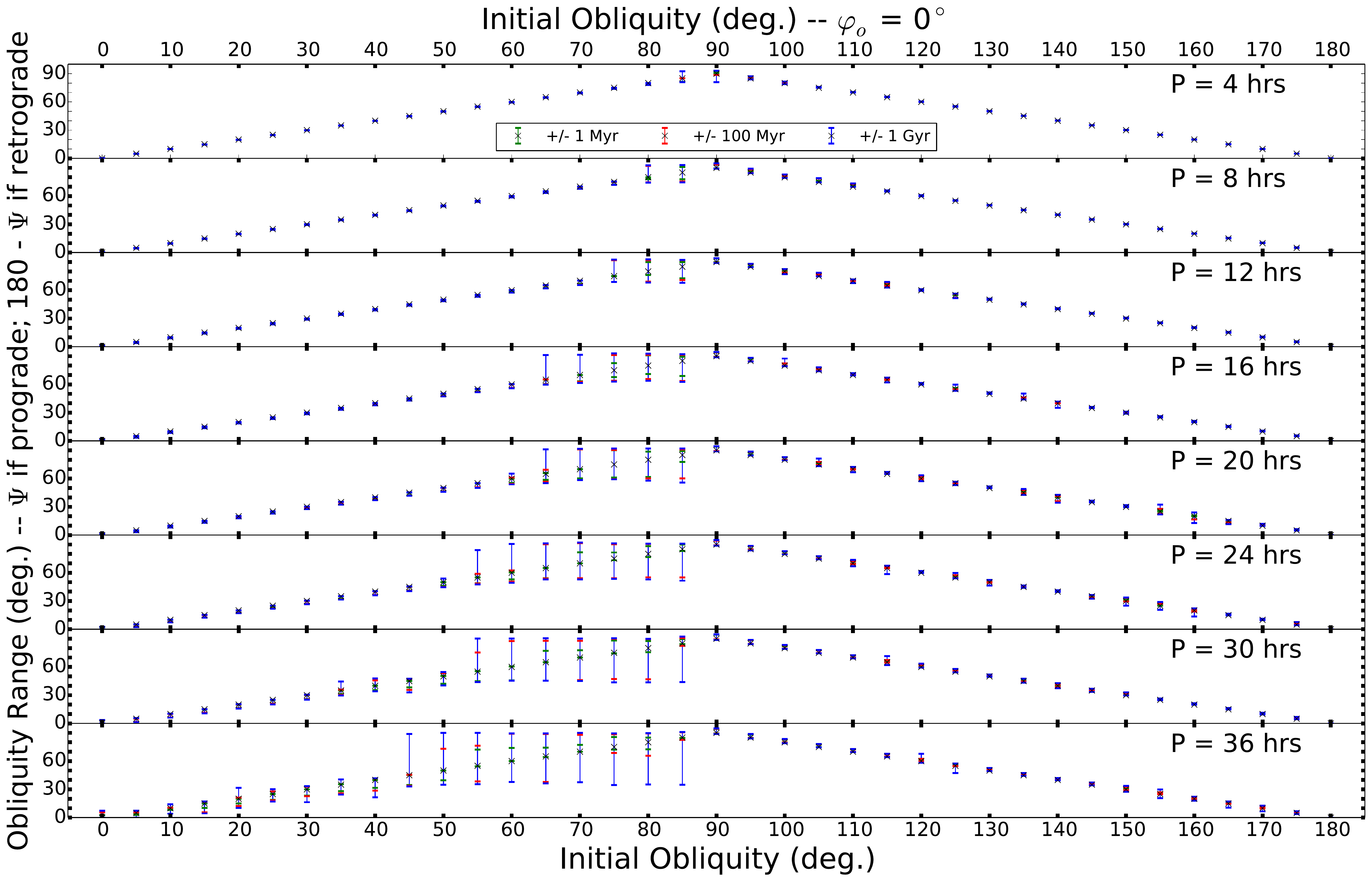}
\caption{Obliquity variation of a hypothetical young Venus considering for
various initial obliquities ($\Psi$) and rotation periods (P).  We calculate the
variations for two different initial azimuths:  $\varphi_0=0^\circ$ is shown
here, and $\varphi_0=180^\circ$ is shown in Figure \ref{figure:obl_var2}.  
The colored bars indicate the range of obliquity variation
over $\pm1$ Myr (green), $\pm100$ Myr (red), and $\pm1$ Gyr (blue).  Obliquities greater than
90$^\circ$ are considered to spin in retrograde and those less than 90$^\circ$
are prograde relative to the orbital motion.  The largest variations occur for
high prograde initial obliquity, with the larger variations extending to lower
initial obliquity for slower rotation.  Some initially high 
prograde obliquities
obtain retrograde rotation, but they do so temporarily, with a maximum obliquity
of $\sim95^\circ$.}
\label{figure:obl_var1}
\end{figure*}

\begin{figure*}[tbh]
\centering
\includegraphics[width=\linewidth]{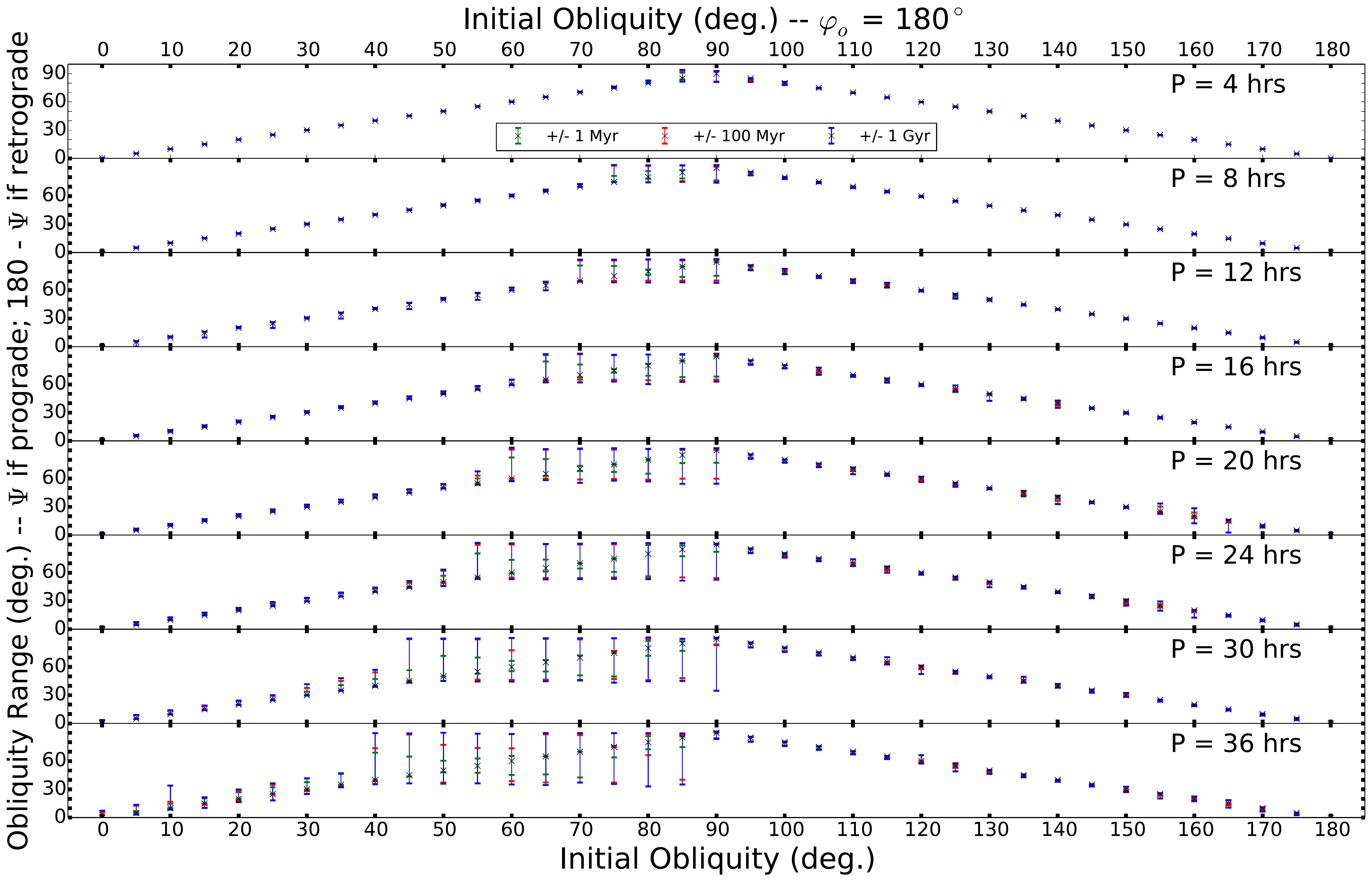}
\caption{Same as Figure \ref{figure:obl_var1}, but for $\varphi=180^\circ$.}
\label{figure:obl_var2}
\end{figure*}

\begin{figure*}[tbh]
\centering
\includegraphics[width=\linewidth]{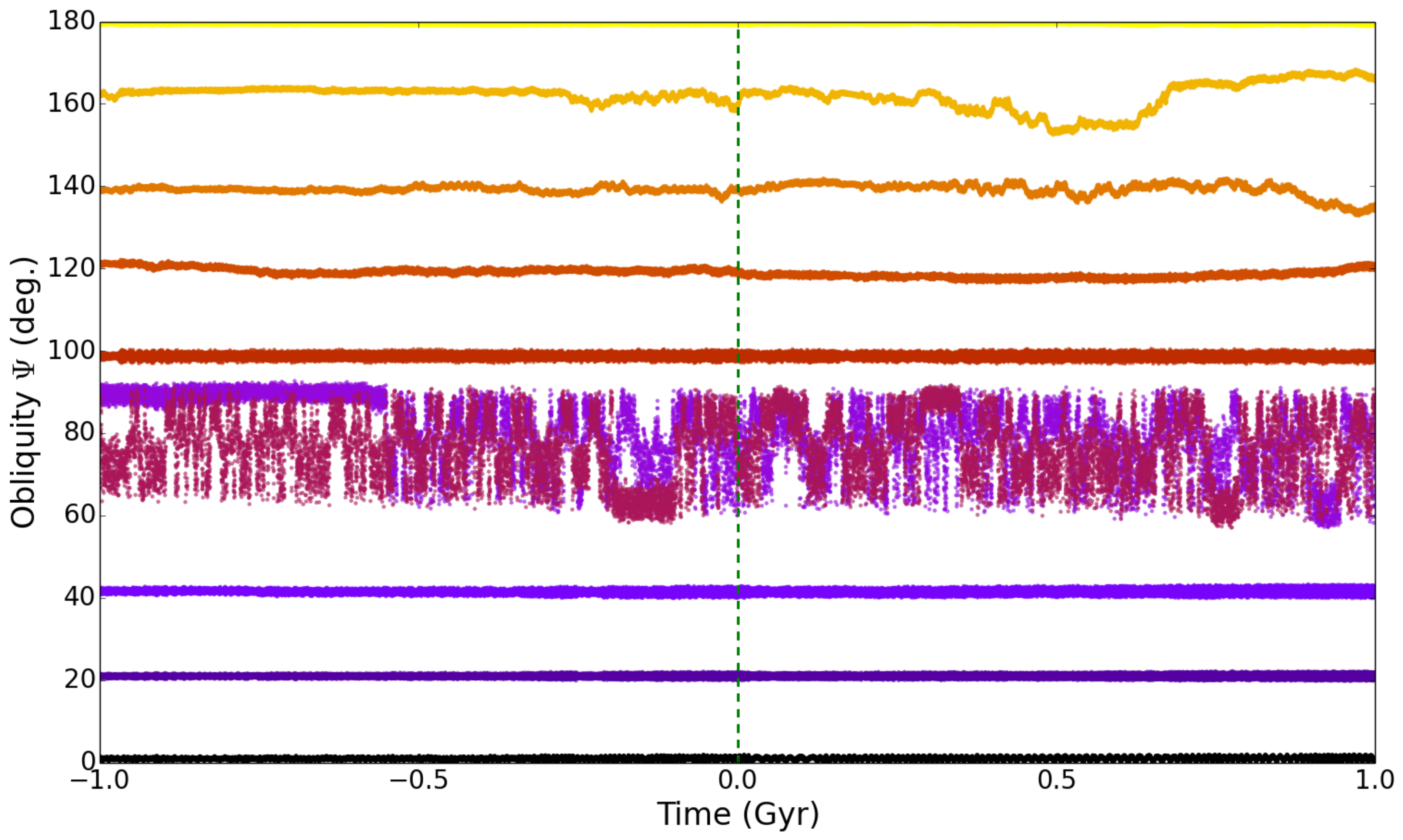}
\caption{Variations of hypothetical Venus's obliquity 1 Gyr into the future and
past for a range of initial obliquities with an initial 20 hour rotation period
and initial azimuth of 180$^\circ$.  The initial obliquities are from
$0^\circ-180^\circ$ for increments of 20$^\circ$ and have been color-coded to
distinguish between overlapping regions.}
\label{figure:obl_ts1}
\end{figure*}

\subsection{Orbital Angular Momentum}

In a single-planet system, neglecting tidal effects and those of stellar
oblateness, a planet's rotational axis would merrily precess around in azimuth
$\varphi$ at a constant rate, but its obliquity $\Psi$ would never change
because the orbit plane would remain fixed.  Thus an important mechanism for
altering planetary obliquity involves the evolution of the \emph{orbit}. 
Because obliquity is the relative angle between the rotation axis and the orbit
normal, either changes in the direction that the axis points in space \emph{or}
changes in the direction of the orbit normal can each alter obliquity
\citep[see, for instance,][Figure 1]{2014AsBio..14..277A}.

We illustrate the evolution of the orbit planes of both Venus and Earth from an
analytical, secular calculation in Figure \ref{figure:VE_pq}.  Figure
\ref{figure:VE_pq} shows the variations in direction of the orbital angular
momentum vectors over 500,000 years.  The motions are of similar magnitude. 
Because Earth and Venus have similar masses and because they each provide the
primary influence on the orbital evolution of the other
\cite[e.g.,][]{SolarSystemDynamics}, their orbital precessions are qualitatively
similar.  Interestingly, Mercury drives the second most important influence on
both Venus and Earth owing to its high orbital inclination relative to both the
ecliptic (the plane of Earth's orbit) and the invariable plane (the plane of the
net angular momentum of the entire Solar System).

The orbital variations of Venus and Earth involve some changes in the orbital
inclination of the two planets, represented by the distance of the lines in
Figure \ref{figure:VE_pq} from the origin.  The primary effect, though, is
counterclockwise near-circular changes that correspond to the precession of the
orbit through space.  We call that effect nodal precession, as it drives
monotonic increases in the element known as the orbit's ascending node, the
angle at which the planet comes up through the reference plane from below.

We show the effective period of the nodal precession of the orbits of Venus and
Earth in Figure \ref{figure:VE_nodal_period}.  Although the precession rate
changes as the orbit inclinations of each planet vary, the long-term average
precession rate for both planets is in the vicinity of $\sim70,000$ years.

\subsection{Spin Chaos}

Through the integration of a secular solution and frequency analysis,
\citet{1993Natur.361..608L} and \citet{1996CeMDA..64..115L} showed that chaos
can be induced when the axial (spin) precessional frequencies are commensurate
with the secular eigenmodes of the Solar System (the
drivers of nodal precession).  Specifically when the spin precession frequency
crosses the eigenmodes associated with secular frequencies $s$1 -- $s$8
($0-26$\textquotesingle\textquotesingle / yr) that are associated with orbital
variations of the planets (including nodal precession), chaotic obliquity
evolution can result.  As a result of this interaction, the obliquity of our
hypothetical Venus can vary substantially.  Weaker secular frequency eigenmodes
can produce additional chaotic zones, albeit smaller in amplitude and
potentially with a longer timescale to develop \citep[e.g.,
][]{2014ApJ...790...69L}. 

Figure \ref{figure:precession}a illustrates the chaotic zones as a function of
initial obliquity $\Psi_0$ for a hypothetical Venus with a 20-hr rotation
period.  The $y$-axis represents the actual average axial (spin) precession rate
$\dot{\varphi}$ in arcseconds per year.  Positive rates here
correspond to clockwise precession as viewed from above the orbit normal;
negative rates correspond to counterclockwise precession, as occurs for
obliquities $\Psi>90^\circ$ (retrograde rotation).

In general the curve of precession rates in Figure \ref{figure:precession}a varies
as a smooth cosine from $+\alpha$ to $-\alpha$, as expected from Equation
\ref{eq:precession_rate}.  However between
$\sim0$\textquotesingle\textquotesingle/yr~
and $26$\textquotesingle \textquotesingle/yr~ the obliquity becomes chaotic, ranging
freely over this span as a function of time regardless of where in that region
the initial obliquity would place it.  For this rotation rate, the primary
chaotic obliquity zone ranges from $\Psi\sim60^\circ$ to $\Psi=90^\circ$.

The power spectrum of the orbit angular momentum direction vector (like that
shown in Figure \ref{figure:VE_pq}) is shown in Figure
\ref{figure:precession}b.  The peaks in this power spectrum labeled $s$1-$s$8
correspond to known Solar System secular eigenfrequencies that result from the 8
interacting Solar System planets.  The secular eigenfrequencies bracket the
$0$\textquotesingle \textquotesingle/yr-$26$\textquotesingle
\textquotesingle/yr~ chaos
region for the 20-hour-rotation Venus, correlating with the chaotic zones in
Figure \ref{figure:precession}a.  The areas labeled $r$1-$r$4 are clusters of
lower-grade retrograde peaks in the frequency power spectrum that we will
discuss further in Section \ref{section:grid_rotation}.

We show a similar plot for a 24-hour-rotation Moonless Earth in Figure
\ref{figure:precession_earthmoo} for comparison.  The Moonless Earth plot shows
the previously known chaotic regions in $\dot{\varphi}$
space, though their correlation with the secular eigenfrequences is poorer than
the hypothetical 20-hour-rotation Venus case.  \citet{2014ApJ...790...69L}
showed that while the chaotic range of obliquities for a Moonless Earth does
indeed extend from $\Psi=0^\circ$ up to $\Psi\sim85^\circ$, the chaotic behavior
is not uniform throughout that range.  

In fact, \citet{2014ApJ...790...69L} find two separate and independent major
chaotic zones:  one from $\Psi=0^\circ$ to $\Psi=45^\circ$, and one from
$\Psi=65^\circ$ to $\Psi=85^\circ$.  While the region between these two major
zones is also chaotic, it is only weakly chaotic.  That connecting region serves
as a narrow `bridge' across which it is possible for planets to traverse, though
only  with substantially reduced probability \citep{2014ApJ...790...69L}.

\begin{figure*}[tbh]
\centering
\includegraphics[width=\linewidth]{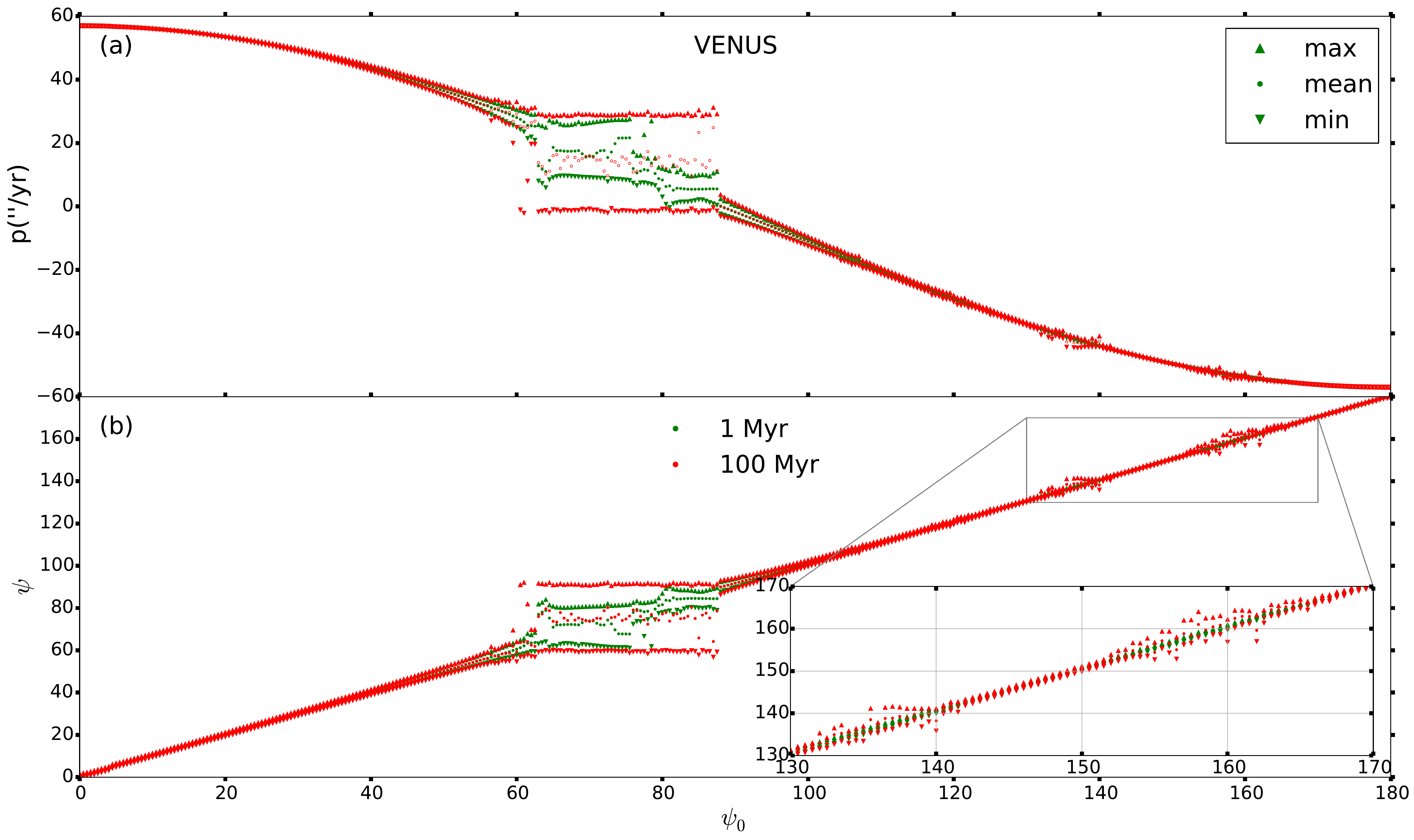}
\caption{Maximum, mean, and minimum variations designated by the up-triangle,
filled circle, and down triangle, respectively, in (a) the precession constant
and (b) the obliquity for a hypothetical Venus with a 20 hr rotation period. 
The points are color coded to signify how the parameters change over 1 Myr
(green) and 100 Myr (red).  The inset in (b) shows potential chaotic zones that
appear to be present in small ranges of retrograde obliquity.}
\label{figure:Chaos_zones}
\end{figure*}    

\begin{figure*}[tbh]
\centering
\includegraphics[width=\linewidth]{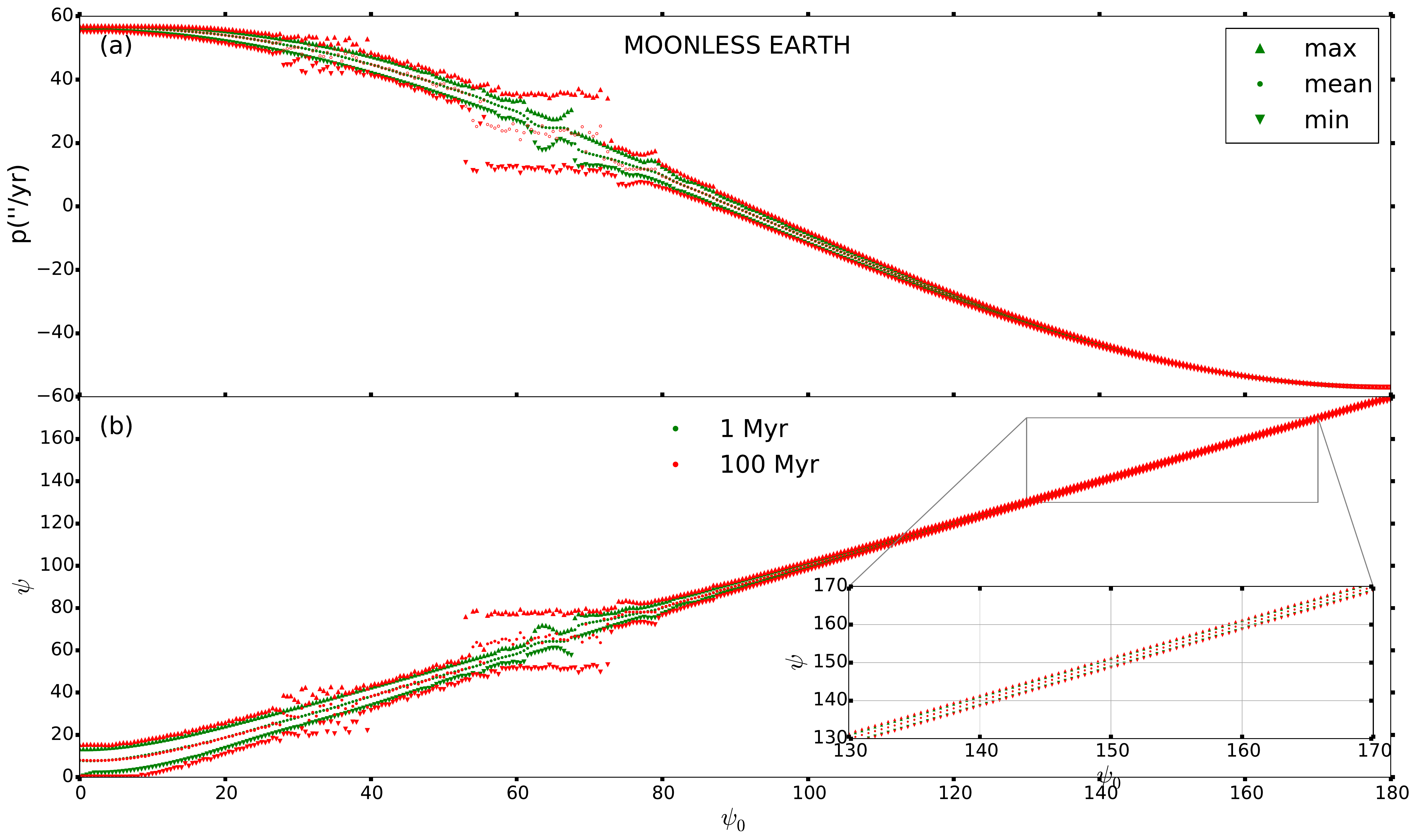}
\caption{Similar to Figure \ref{figure:Chaos_zones}, here we plot similar
maximum, mean, and minimum extents of variation in obliquity $\Psi$, but this
time for a 24-hour-rotation-period Moonless Earth.  This plot verifies the lower-resolution integrations
presented in \citet{moonless.Earth} in that, unlike for hypothetical early
Venus, no chaotic obliquity variations are
evident in any of the retrograde Moonless Earth cases.}
\label{figure:EARTHMOO_chaos}
\end{figure*}

\section{Numerical Results}\label{section:results}

\subsection{Coarse Grid}

We initially explore the obliquity of hypothetical Early Venuses by numerically
integrating the obliquity variations forward to +1~Gyr and backward to -1~Gyr
over a coarse grid of rotation rates and initial obliquities.  We show a summary
of the resulting obliquity variations as a function of initial obliquity and
rotation rate in Figures \ref{figure:obl_var1} and \ref{figure:obl_var2}.   The
difference between the two figures is the azimuthal direction in which the
rotation axis initially points, which effectively corresponds to where the
planet is in its rotation axis precession (i.e., the precession of the
equinoxes).

\begin{figure*}[tbh]
\centering
\includegraphics[width=\linewidth]{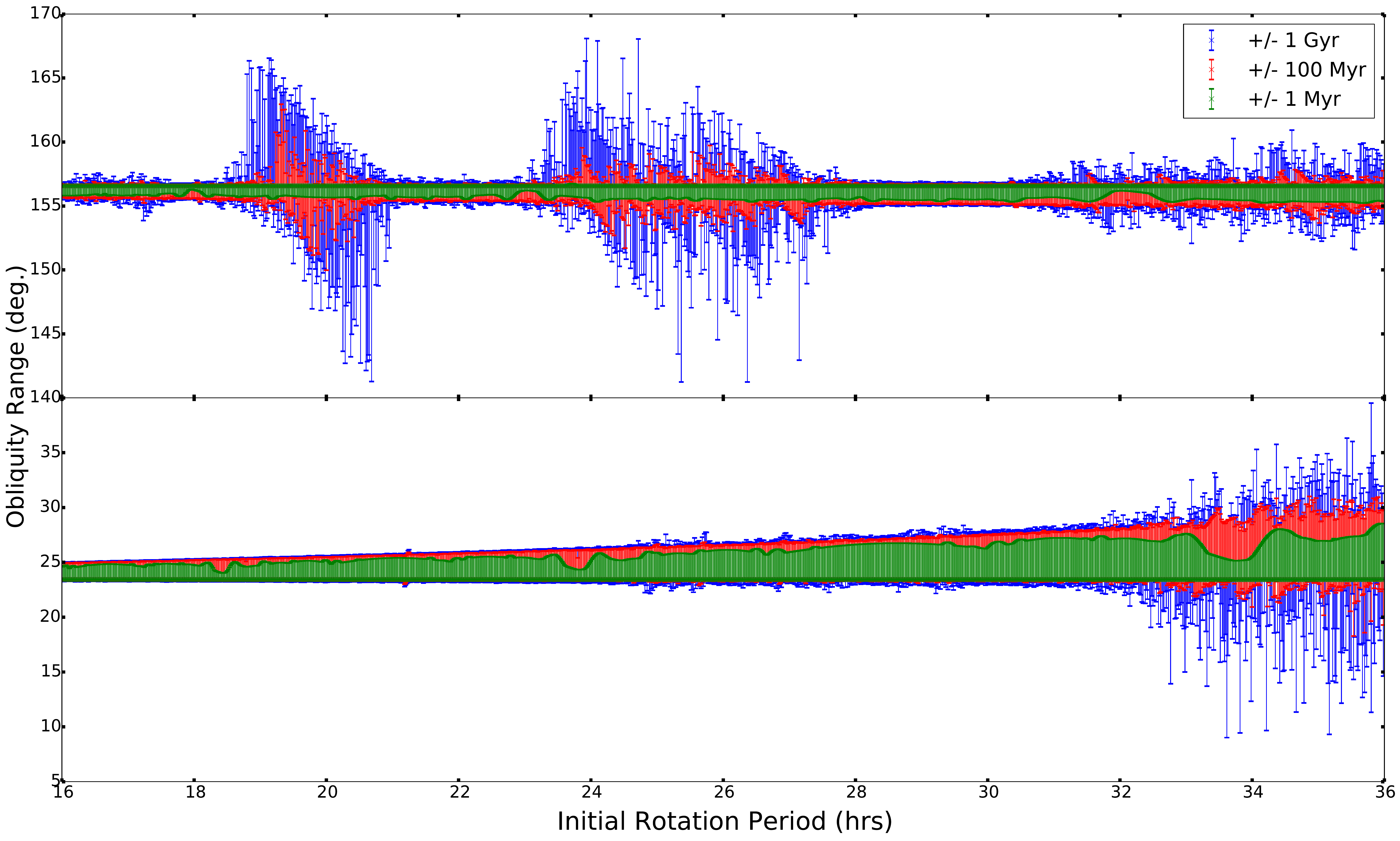}
\caption{Variations of hypothetical Venus's obliquity for an initially
Earth-like obliquity ($\Psi=23.45^\circ$) for retrograde (top panel) and
prograde (bottom) rotations.  The initial rotation periods are varied in 1
minute increments from 16 hr to 36 hr.}
\label{figure:Per_var}
\end{figure*} 

We consider the results in the context of the values for the precession constant
$\alpha$ shown in Figure \ref{figure:Spin_param}.  A rapid Venus spin period of
4 hours drives a considerably large equatorial bulge (oblateness and $J_2$), which
in turn leads to a high precession constant of
335\textquotesingle\textquotesingle/yr.  As this value greatly exceeds
any of the frequencies of significant power in the orbital angular momentum
direction power spectrum (Figure \ref{figure:precession}b), nearly all of the
resulting obliquity variations remain within tight ranges ($\pm\sim2^\circ$,
similar to present-day Earth obliquity variability with the Moon) and
non-chaotic.  

At very high obliquity, however, near-resonant conditions can occur due to the
$\cos{\Psi}$ dependence in Equation \ref{eq:precession_rate} for rotation axis
precession.  Hence for initial conditions with $\Psi_0=85^\circ$ and
$\Psi_0=90^\circ$ we see moderately variable and chaotic obliquity
variations, even for this fast 4-hour rotation period.

Retrograde rotations for the 4-hour rotation period ($\Psi>90^\circ$) show very
low variability.  Similarly small variations were seen for retrograde Moonless
Earths \citep{moonless.Earth}.

Proceeding to slower rotation rates of 8, 12, and 16 hours, the low-obliquity
end of the chaotic region drops to $\Psi=75^\circ$,  $\Psi=70^\circ$, and
$\Psi=65^\circ$ respectively for initial axis azimuth of $\varphi_0=180^\circ$ in
Figure \ref{figure:obl_var2} (with similar results at $\varphi_0=0^\circ$ in
Figure \ref{figure:obl_var1}).  This downward expansion of the chaotic zone is
consistent with the effects of lower obliquity on precession rate from Equation
\ref{eq:precession_rate}.  As the slower rotation reduces the planet's $J_2$, it
also diminishes its precession constant $\alpha$.  Hence a lower obliquity value
$\Psi$ can result in similar rotation axis precession rates as the
high-obliquity 4-hour rotation case.

Importantly the slower rotation rate does not introduce new chaotic regions at
lower obliquity, but rather slightly reduces the maximum obliquity at the top of
the chaotic zone and substantially reduces the minimum obliquity at the bottom
of the zone, leading to a wider zone overall.  Hypothetical
Venuses that start anywhere within the chaotic region have their obliquities
vary across the entire range from the lower limit to near $90^\circ$ over a Gyr.

These same trends continue as we proceed down Figures \ref{figure:obl_var1} and
\ref{figure:obl_var2} to longer rotational periods.  From 20-hr up through 36-hr
rotation periods, the overall extent of the chaotic zone at high obliquities
grows.  The upper limit of the primary chaotic region is always above
$\Psi\sim85^\circ$, but the lower boundary extends all the way down to below
$\Psi=45^\circ$ for a 36-hour siderial rotation.

Interestingly, a new, more weakly chaotic region also appears at slower rotation
rates.  At 20, 24, 30, and 36 hours, some smaller initial obliquities $\Psi_0$
below the edge of the primary chaotic zone show moderately variable
obliquities.  When the initial obliquity is $\Phi_0=50^\circ$ in the 24-hour
rotation case at $\varphi_0=180^\circ$ (Figure \ref{figure:obl_var2}), for
instance, the obliquity varies in the range $45^\circ \leq \Phi \leq 60^\circ$
over $\pm1$~Gyr.

In the 30-hour and 36-hour period cases, this lower-obliquity weakly chaotic
zone grows.  At 36-hours, it includes all of the initial obliquities smaller
than the primary zone, from $0^\circ \leq \Phi \leq 35^\circ$.  Hypothetical
Venuses with initial obliquities inside this weaker zone show increased
obliquity variability at the $\pm\sim15^\circ$ level.  But with the exception of
the $\Psi_0=10^\circ$, $\varphi_0=180^\circ$ case the total $\pm1$~Gyr
variability does not encompass the entire extent of the weaker chaos zone. 
These cases only show moderately increased obliquity variability, similar to
that of Moonless Earths, which have broadly comparable nodal and axial
precession rates.

Although rapidly-rotating retrograde Early Venuses lack the large-scale
variations found for high prograde obliquities, in some simulations their
obliquities vary much more than those of Earth would have if it lacked a
large moon and rotated in the retrograde sense.  In the 20-hour-rotation,
$\varphi_0=180^\circ$ case, for instance (Figure \ref{figure:obl_var2}),
obliquity variations of similar magnitudes to those in the weakly chaotic
low-obliquity regime appear at $\Psi_0=155$, $\Psi_0=160$, and $\Psi_0=165$ for
instance.  We did not expect to find chaotic obliquity behavior for retrograde
rotations given the high degree of stability found in retrograde Moonless Earths
\citep{moonless.Earth}.

\subsection{Closer Look at Venus with a 20-Hour Rotation
Period}\label{section:20hr}

\subsubsection{Time-Series}

Focusing on the results for Venus with a 20 hr rotation period, which show
unexpected chaos for some retrograde obliquities, Figure \ref{figure:obl_ts1}
shows the full $\pm1$~Gyr time histories for the obliquity $\Psi$ of
hypothetical Venuses with initial obliquities $\Psi_0$ spaced out every
$20^\circ$. The low initial obliquity cases $\Psi_0=0^\circ$, $20^\circ$, and
$40^\circ$, have obliquities that vary within narrow ranges and show no chaotic
long-term behavior.  Similarly, the $\Psi_0=100^\circ$ and $\Psi_0=180^\circ$
cases vary uniformly within a tight band with no chaotic behavior on either the
medium- or long-term.

The $\Psi_0=60^\circ$ and $80^\circ$ cases are within the primary chaotic
region.  These two cases bounce around between three smaller chaotic subregions
(except for the $\Psi_0=80^\circ$ case which manages to find its way out of the
chaotic region beyond 550~Myr in the past).

The retrograde $\Psi_0=120^\circ$, $\Psi_0=140^\circ$, and $\Psi_0=160^\circ$
cases display behavior qualitatively different from any seen in the Moonless
Earth case, on the other hand.  In these cases, our hypothetical Venus'
obliquity varies within a relatively tight $\pm2^\circ$ band on both short- and
medium-term timescales.  On longer timescales approaching 10-100~Myr, however,
the center of that tight band wanders around in obliquity $\Psi$ space up to
$\pm10^\circ$ (in the $\Psi_0=140^\circ$ and $\Psi_0=160^\circ$ cases; the
$\Psi_0=120^\circ$ case is less adventurous).

These odd retrograde cases and the chaotic $\Psi_0=60^\circ$ and
$\Psi_0=80^\circ$ situation are distinct.  In the primary chaotic zone obliquity
varies within a single chaotic subregion while periodically and vary rapidly
traversing wide chaotic `bridges' \citep{2014ApJ...790...69L} to neighboring
chaotic subregions.  These transitions between subregions last for only of order
a single precession period, or $\sim70,000$~years for hypothetical Early Venus. 
In contrast, the retrograde rotators with $\Psi_0=140^\circ$ and
$\Psi_0=160^\circ$ continue rapid, short-term variations on $10^5$-year
timescales.  But they slowly vary in obliquity on $10^7$-year timescales
instead of nearly instantaneous alteration of their variations into a new regime
as in the primary chaotic zone.  

A glance back at Figure \ref{figure:precession} shows that the unexpected
retrograde long-term variability corresponds to the locations of weak
frequencies in Venus' orbit variations.  However, the power associated with
those peaks is a factor of $10^6$ lower than the primary Solar
System $s$ eigenfrequencies.  Furthermore, in the 24-hour Moonless Earth
case shown in Figure \ref{figure:precession_earthmoo} for comparison, similar
peaks in Earth's orbital angular momentum direction frequency do not yield
corresponding variability for retrograde rotators.  Similarly, not all
hypothetical Venuses show this behavior; retrograde rotators
with 4 and 8 hour periods show little long-term obliquity variation.

\subsubsection{Fine Grid in Obliquity}

To further investigate this unexpected chaotic obliquity evolution in retrograde
rotators, we ran additional obliquity variation integrations at very high
resolution in initial obliquity $\Psi_0$.  In this section we analyze the
20-hour-rotation-period Early Venus specifically because it shows the strongest
anomalous retrograde variability.  Figure \ref{figure:Chaos_zones} shows our
results using a grid of 361 different $\Psi_0$ values spaced out every
$0.5^\circ$.  In this portion of the investigation we elect to integrate out only
to $\pm100$~Myr to allow for improved resolution in $\Psi_0$ within our available
computing power.

We show the results of this fine $\Psi_0$ integration in Figure
\ref{figure:Chaos_zones}.  We also show the analogous plot for
Moonless Earths in Figure \ref{figure:EARTHMOO_chaos}, seeing as
\citet{moonless.Earth} did not perform such a high-resolution grid of
simulations.  

Similar to the coarser-gridded \citet{moonless.Earth} result, the Moonless Earth
shows moderately wide $\pm\sim10^\circ$ obliquity variations from
$\Psi_0=0^\circ$ through $\Psi_0=55^\circ$ over $\pm100$~Myr.  More distinct
chaotic subregions then extend up to $\Psi_0=85^\circ$.  Even when viewed at
this high resolution, though, the Moonless Earth shows no signs of anomalous
behavior for retrograde rotations (although in retrospect Figure 10 from
\citet{moonless.Earth} may show the incipient onset of such variations at 1~Gyr
timescales).

The 20-hour hypothetical Venus in Figure \ref{figure:Chaos_zones} shows very
tight obliquity ranges from $\Psi_0=0^\circ$ through $\Psi_0\sim55^\circ$ or so,
followed by the single large primary chaotic region from $\Psi_0=60^\circ$ to
$\Psi_0=90^\circ$ as discussed above.  The smaller chaotic subregions reveal
themselves when looking at shorter 1~Myr timescales (green).

Although hypothetical 20-hour Venus shows a somewhat simpler chaotic obliquity
variation structure than Moonless Earth for prograde initial conditions, the
opposite is true once the planets flip over into regrograde rotation at
$\Psi_0>90^\circ$.  At retrograde obliquities the Moonless Earth shows
minimal obliquity variations even over 100~Myr timescales.

Retrograde 20-hour Venus, on the other hand, shows broadly stable 
obliquities but with four modestly more variable regions centered around 
$\Psi_0=100^\circ$, $\Psi_0=120^\circ$, $\Psi_0=135^\circ$, and 
$\Psi_0=158^\circ$.  These regions seem to coincide with the low-power 
peaks in orbital frequency space shown in Figure 
\ref{figure:precession}.  However, we do not at present understand why 
these peaks are important for Venus but not for Moonless Earths, which 
show similar peaks in orbital frequency space.

\subsubsection{Fine Grid in Rotation Period}\label{section:grid_rotation}

For one last exploration of this unexpected retrograde behavior we do another
set of integrations at high resolution, but this time in rotation-period space. 
Starting with $\Psi_0=23.45^\circ$ and $\Psi_0=180^\circ-23.45^\circ$, we show
obliquity variations as a function of rotation rate in high resolution in Figure
\ref{figure:Per_var}.  This plot shows the obliquity variations for hypothetical
Venuses over three timescales:  $\pm1$~Myr (green), $\pm100$~Myr (red), and
$\pm1$~Gyr (blue).

For a retrograde Earth-like obliquity of
$\Psi_0=156.55^\circ=180^\circ-23.45^\circ$, three independent wide-variability
regions occur at rotation periods $P_\mathrm{rot}$ of 18.8-21 hours, 23.5-28
hours, and 31.5-36+ hours.  These correspond respectively to the $r4$, $r3$, and
$r2$ retrograde precession frequencies from Figure \ref{figure:Chaos_zones}b. 
Presumably another similar region exists for even longer rotation periods
corresponding to $r1$.

All told, while unexpected, these chaotic zones at retrograde rotations only
show modest total variability --- $\pm7^\circ$ in the worst case over 1~Gyr. 
Understanding their origin is important for evaluating the suggestion of
\citet{moonless.Earth} that ``if initial planetary rotational axis orientations
are isotropic, then half of all moonless extrasolar planets would be retrograde
rotators, and these planets should experience obliquity stability similar to
that of our own Earth, as stabilized by the presence of the Moon."  While our
results show that the most variable retrograde hypothetical Venuses are more
stable than the standard Moonless Earths, the same may not be true for
retrograde-rotating planets in all cases.

\section{Conclusions}\label{section:conclusions}

We investigate the variations in obliquity that would be expected for
hypothetical rapidly rotating Venus from early in Solar System history.  These
hypothetical Early Venuses allow us to investigate the conditions under which
Venus' climate may have been sufficiently stable as to allow for habitability
under a faint young Sun.  

Additionally, they also serve as a comparitor for potentially habitable
terrestrial planets in extrasolar systems.  While previous work on a moonless
Earth effectively modeled a single point of comparison, the present work
provides a second comparator from which we can start to imagine a more general
result.  These intensive, single-planet studies complement those of generalized
systems \citep{2007Icar..188....1A}.

We show that while retrograde-rotating hypothetical Venuses show short- and
medium- term obliquity stability, an unusual and as-yet-understood long-term
interaction drives variability of up to $\pm7^\circ$ over Gyr timescales.

The very low variability of low-obliquity hypothetical Venuses over a range of
rotation rates provides additional evidence that massive moons are not
\emph{necessary} to mute obliquity variability on habitable worlds.  We show
that even in the Solar System the increased rotational axis precession rate
driven by Venus' closer proximity to the Sun is sufficient to push Venus into a
benign obliquity variability regime.  Indeed, Figure \ref{figure:precession}
for example indicates that for present-Earth-like initial obliquities
($\Psi_0=23.45^\circ$), the overall obliquity variability over 100~Myr for Venus
with a 20-hour rotation period is similar to that for the real Earth
with the Moon.

More rapid rotational axis precession will naturally result on planets in the
habitable zones of lower-mass stars.  While these stars' gravity is
proportionally lower, their disproportionately fainter luminosities drive the
habitable zone inward from that around the present-day Sun.  Thus, for similar
orbital driving frequencies -- i.e., a clone of the Solar System around 
a lower-mass star -- stellar 
gravity alone would be sufficient to
push a habitable planet's obliquity variations into a benign regime.

Of course tides provide a drawback to using stellar proximity to speed 
rotational precession.  In any real system, in addition to the obliquity 
variations that we describe here, tides will simultaneously act to 
upright a planet's rotation axis and slow its rotation rate Tidal 
effects will be even more important on habitable zone planets around 
lower mass stars than they are for Earth and Venus around the Sun.

Hence a potential avenue for future work will be to couple the adiabatic 
obliquity variations that we describe here to tidal dissipation over 
time. Given that the natural variability within chaotic zones is much 
more rapid than tidal timescales, we suspect that the primary effect of 
tides will be through rotational braking.  A planet with slowing 
rotation could traverse through various obliquity behavior regimes over 
its lifetime.  Such a planet might then potentially have multiple 
possible interesting and chaotic pathways toward tidal locking, as 
opposed to the simpler slow obliquity reduction that would be expected 
in a 1-planet system.

\acknowledgements  The authors acknowledge support from the NASA Exobiology
Program, grant \#NNX14AK31G.

\bibliographystyle{apj}
\bibliography{references}


\end{document}